\documentclass[12pt]{article}
\usepackage[top = 2 cm, bottom = 3 cm, left = 3 cm, right = 1.5 cm]{geometry}
\usepackage{authblk, cite, color, amssymb, amsmath}
\usepackage[colorlinks = true, linkcolor = blue, citecolor = red]{hyperref}
\usepackage[dvipsnames]{xcolor}

\begin{document}

\title{Information-Probabilistic Description of the Universe}

\author{\bf Merab Gogberashvili}
\affil{\small Javakhishvili Tbilisi State University, 3 Chavchavadze Avenue, Tbilisi 0179, Georgia \& \authorcr
Andronikashvili Institute of Physics, 6 Tamarashvili Street, Tbilisi 0177, Georgia}
\maketitle

\begin{abstract}
We describe the universe as a single entangled ensemble of quantum particles. The total entropy of this world ensemble, which can be expressed as a sum of information, thermodynamic and entanglement components, is assumed to be always zero. This condition suggests information quantization, which we associate with the Planck's action. Then the entropy neutrality condition for the universe leads to the zero-action principle. We show that the main concepts of classical space-time and gravity naturally emerge in this picture. A generalized least action principle, which embraces the maximal entropy principles of information theory, is introduced.

\vskip 5mm
PACS numbers: 04.50.Kd, 89.70.Cf, 05.30.Ch, 03.65.Ud
\vskip 2mm
Keywords: Information conservation; Entropy of the universe; Stochastic action
\end{abstract}


\section{Introduction}

Many authors consider the so called emergent theories \cite{Car} in which classical space-time is not a fundamental concept \cite{deBr, Sakh, Bek, BCH, Haw-1} and gravity is defined only for the matter in bulk (like thermodynamics or hydrodynamics) and not for individual elementary particles \cite{Rov-2, Rov-3, Jac, Pan, Ver}. In understanding the origin of gravity, which is far from being complete, and its relations to other interactions in the universe the information-probability methods may play a central role. The key concept in this direction is entropy, which is a powerful tool in thermodynamics and information and quantum theories. Entropy is an additive quantity in general, having distinct ingredients, what allows us to model different aspects of physical systems using a similar mathematical framework.

In classical thermodynamics the Boltzmann's entropy \cite{Stat-1, Stat-2},
\begin{equation} \label{S}
  S = \ln N ~,
\end{equation}
quantify the amount of information that is lacking in order to identify the microstate of a system from $N$ possibilities\footnote{Throughout the paper we measure entropies in nuts and assume the Boltzmann's constant $k_B=1$. To convert the entropy into bits one can perform the standard replacement, $\ln \to \ln 2 \log_2$.}, which are compatible with the given macrostate.

In information theory, entropy is a measure of the uncertainty of information associated with some random variable. In this context, the term usually refers to the Shannon's entropy \cite{Shan},
\begin{equation} \label{I}
  I = - \sum_{i=1}^N p_i \ln p_i ~,
\end{equation}
where $p_i$ denotes the probability that an entity will be the $i$-th.

Entropy in quantum mechanics corresponds to the lack of information about the outcomes of a measurement process. For either a pure state specified by a wave function, or a mixed state specified by a density matrix $\rho$ the amount of (or lack of) information about the elements of a quantum system is usually measured by the von Neumann entropy \cite{Ent-1, Ent-2, Ent-3},
\begin{equation} \label{S-vN}
  {\cal S} = - {\rm tr} \left(\rho \ln \rho\right) ~.
\end{equation}
This kind of entropy captures both classical and quantum uncertainty, i.e. it contains additional information concerning the quantum entanglement, which is not present in the density matrices of subsystems. A quantitative description of this quantum property, the entanglement entropy, is a form of information entropy that describes needed information to have a full knowledge about other parts of the quantum system. The entanglement entropy is being used for a wide spectrum of models like quantum information processing \cite{Ent-2, Ent-3}, quantum phase transitions \cite{Phas-Tr-1, Phas-Tr-2} and black holes \cite{BH-1, BH-2, BH-3}. It is different from the standard statistical entropy (\ref{S}) in many respects, but several characteristic thermodynamic relations remain valid \cite{H=S-1, H=S-2}. In some models the entanglement entropy even is found to be divergent \cite{S-div-1, S-div-2, S-div-3}.

Entropy in physical theories quantifies two seemingly opposite entities: uncertainty (or ignorance) and information (or order). However, there are deep reasons to expect relations between different kinds of entropies \cite{Schl}, since using information an observer can manipulate physical systems and change their thermodynamic or quantum entropies. For instance, the Shannon's entropy (\ref{I}) reduces to the Boltzmann's one (\ref{S}) when an observer assumes the equal probabilities for each macrostate,
\begin{equation}
  p_i = \frac 1N ~.
\end{equation}
It is known that the Shannon entropy of the ensemble is strictly greater than the von Neumann entropy whenever the states in the ensemble are non-orthogonal \cite{Wilde}. So for quantum systems we have the relation,
\begin{equation} \label{S<I}
  I \geq {\cal S} ~,
\end{equation}
that means that there is more uncertainty in a single observable than in the whole of the quantum state described by a density matrix. The equality in (\ref{S<I}), or the reduction of von Neumann's entropy (\ref{S-vN}) to the Shannon's one (\ref{I}), holds for diagonal density matrices.

In general, identification of information and thermodynamic entropies is not valid, $I$ is a subjectively defined quantity, and it is meaningless without an observer for which the probabilities $p_i$ are defined. While $S$ is an objective, an observer independent quantity. A physical system has a fixed value of $S$, whether one knows identity of the microstate or not. When the observer knows in which microstate the system is, $I$ becomes zero (all $p_i$-s in (\ref{I}), except of one, are zero), but $S$ is unchanged. So the concept of entropy in physics is far from being well understood. An example is the case of black holes, where we are not able to associate the entropy with a counting of microstates. In addition, as a consequence of the Unruh effect, it appears that the number of microstates might be related to the motion of the observer. This suggests that a more general definition of entropy than introduced by Boltzmann, Shannon or von Neumann might exist. Examples of the alternative entropy concepts are the R\'{e}nyi and Tsallis entropies, relative entropy, entropy density, etc. \cite{Ent-1, Ent-2, Ent-3, Wilde}. We can mention two reasons that entropy to be a mysterious concept: (i) the contradiction between the time-irreversible second law of thermodynamics and the time-reversal invariant laws of microphysics and (ii) the fact that the process of coarse-graining by which one defines which states are indistinguishable is partially subjective.

We can strictly relate the information and thermodynamic entropies with each other by adopting the principle of information-entropy conservation for any closed system, be it classical or quantum \cite{I-cons}. This principle is true for classical systems, due to the conservation of phase space volume in Hamiltonian dynamics, when a distribution function evolves according to the Liouville theorem. In the case of information theory, the total entropy of the information coding device plus its environment is conserved, since the thermodynamic entropy for the environment increases after the erasure of information. Another example concerns the famous Maxwell demon paradox with the violation of second law if one ignores demon's information and where the entropy remains conserved only for the complete system of the demon, the gas and the environment. The fact that information is physical takes on a more demanding role for quantum case, where information can neither be cloned \cite{I-clon} nor deleted \cite{I-del}. In quantum teleportation process the information conservation manifests itself when a reconstruction of the unknown initial state at the receiver requires a classical communication channel linking to the sender, and the information transfer is limited by the speed of light \cite{Teleport}.

If we adopt the principle of information-entropy conservation we should conclude that due to the second law of thermodynamics (that the statistical entropy always increases) in standard definitions of total entropy of a physical system some negative component is missing. While the usual definitions of entropies, like (\ref{S}), (\ref{I}) and (\ref{S-vN}), are non-negative, it appeared that the conditional entanglement entropy,
\begin{equation} \label{H}
  H = {\cal S}_{tot} - {\cal S}~,
\end{equation}
where ${\cal S}_{tot}$ denote the joint entropy of quantum states and ${\cal S}$ is the marginal von Neumann entropy of subsystems, can be negative \cite{Wilde, Neg-S-1, Neg-S-2, Neg-S-3, Neg-S-4, Neg-S-5, Neg-S-6, Neg-S-7, Neg-S-8}. This kind of entropy is known as the useful tool for quantitative measure of the entanglement of a quantum system with many degrees of freedom \cite{Ent-S-1, Ent-S-2}. The examples with negative $H$ are the pure entangled quantum states, which have stronger than classical spatial correlations, when one can be more certain about the joint state of a quantum system than about any of its individual parts. Then in complex models, which use combinations of different types of entropy, confusions with the sign can arise. An example is the von Neumann's multi-component measurement scheme, in which the quantum system, memory (or apparatus), and the observer himself are involved, and the entropy component for memory yields a negative value. There have been a number of notable interpretations of the negative entanglement entropy, as a potential information \cite{HOW}, a reversible work \cite{RARDV}, or a memory filling the system similar to the Dirac's sea \cite{Song}.

The fact that the joint quantum entropy ${\cal S}_{tot}$ in (\ref{H}) can be less than a sum of marginal entropies of subsystems, ${\cal S}$, is one of the most fundamental differences between the classical and quantum information, and it is a key observation in quantitative realization of the principle of information-entropy conservation used in informational interpretation of quantum mechanics \cite{Cerf-Adami,Zei-1, Zei-2}. This principle allows creation of entropy in the measurement device which is counterbalanced by the negative entropy of the quantum system itself, and results in the conservation of the total entropy in the measurement process. In informational models the collapse of the wavefunction does not occur and the quantum probabilities arise in the unitary time development of the measurement thanks to the negative entropy of the 'unobserved' quantum system. For example, the decrease of entropy during the black hole evaporation is exactly balanced by the increase of entropy in the emitted radiation, so that the complete process of Hawking radiation in this model is unitary.

It is known that 'classical' definitions of entropy lead to problems, when one tries to apply them to the whole universe. If the universe at the early stages was in equilibrium (as the CMB appears to suggest), already having a maximal $S$, its entropy could not have increased over its lifetime. Therefore, the initial value of the 'universal' entropy should be close to zero, as this is in the case of von Neumann's entropy for a single universe model (with no observers outside), which is in a huge pure state. This is in contradiction with the fact that thermodynamic entropy for the universe is enormous. The significance of this paradox was pointed out long ago \cite{Penrose} and still there is little consensus about how to define the maximal and minimal values, and the evolution rate of the entropy for the universe \cite{Lin-Ega}.


\section{Zero-entropy principle}

In our opinion, a satisfactory model of the entropy for the universe can be obtained using the ideas of information interpretation of quantum mechanics \cite{Cerf-Adami,Zei-1, Zei-2}, in which a measurement is considered as the interaction of three systems: the quantum object, memory (measurement device) and observer. Then the total entropy of the universe can be assumed to be zero (as it is suggested by von Neumann's model) and, formally, it can be written as the sum of the information, statistical (thermodynamic) and quantum (entanglement) components \cite{Hwan,Song},
\begin{equation} \label{I+S+H}
  I + S + H = 0~.
\end{equation}
This relation is the special case of Schmidt's decomposition for the entangled composite system in pure state and it follows from Araki-Lieb inequalities \cite{Araki-Lieb}. According to the information interpretation, the universe always remains in pure state, i.e. it obeys (\ref{I+S+H}) due to the balancing of the randomness in the $I$-$S$ mixed state of the observer and the measurement device with the negative conditional entanglement entropy (\ref{H}) with the rest of the universe. Since the model only allows a unitary time-evolution, the 'universal' entropy (\ref{I+S+H}) remains zero at all stages of the universe's evolution, while any subsystem has non-zero entropy.

The condition of entropy neutrality of the universe (\ref{I+S+H}) suggests the information quantization \cite{Zei-1,Zei-2}, such as the discreteness of the charge might be thought as the consequence of the validity of Gauss' law for the electrically neutral finite universe. Since information is physical, the existence of the unit of information \cite{Cerf-Adami} should be associated with some physical parameter, like energy or mass \cite{Info-1, Info-2, Info-3}, according to the Landauer \cite{Land-1, Land-2} and Brillouin \cite{Bril} principles, respectively. However, these parameters cannot serve as adequate measures of information, since it is possible for a system to lose information stored in the degenerate states without losing energy \cite{Unr, JMK}. Also, there is no unique standard of mass and energy, which is additive as the electric charge. Moreover, mass and energy have no polarity and usually are taken to be positive.

In our opinion the convenient physical parameter to measure information is the action $A$, which is an additive quantity like entropy, and it also contains positive and negative components and exhibits the unique discrete value, the action quantum $\hbar$ \cite{Ann}. Thus we assume that thermodynamic component of entropy is proportional to the classical action of the system,
\begin{equation} \label{S=A/h}
  S \sim \frac{A}{\hbar}~.
\end{equation}
This relation translates the condition of entropy neutrality of the universe (\ref{I+S+H}) into the null action principle - the sum of all components of the action for a physical system (including the boundary terms) is zero. The consequences of the null-action principle are the appearance of different conservation laws in physics, be it discrete or continuous, and the zero-energy condition for the whole universe, which implies that the universe can emerge without violation of the energy conservation \cite{Fey, Haw-2}. The assumption (\ref{S=A/h}) also means that the minimal information, which is encoded in the response of the universe to the transfer of the elementary action,
\begin{equation}
  A_{min} = \hbar ~,
\end{equation}
describes the value of the Boltzmann's constant, which is presents in (\ref{S=A/h}) as $k_B =1$.

We want to show that the main concepts of classical space-time and gravity naturally emerge within the information-probabilistic model of the universe considered as the unified ensemble of entangled quantum particles \cite{Gog-1, Gog-2, Gog-3}.


\section{Time arrow}

The emergence of the direction of time in our model is a consequence of the second law of thermodynamics written for a quasi-isolated system of particles with
\begin{equation} \label{H=C}
  H = const ~.
\end{equation}
The system can be in $k$ different configurations with the probabilities
\begin{equation}
  P_k \sim e^{S_k} ~,
\end{equation}
where $S_k$ are the conditional thermodynamic entropies of these configurations. We can assign an ordering of configurations in a natural way, since there are no two of them with the same weight, and the configuration labeled by $k_1$ contains the configuration $k_2$ if
\begin{equation}
  S_{k_1} > S_{k_2} ~.
\end{equation}
Therefore, the entropy of the configuration $S_k$ can play the role of a time parameter and the maximum thermodynamic entropy principle \cite{Jay} incorporates an intrinsic directionality in this entropic time: the maximal probability to be observed has a configuration with the largest $S_k$ corresponding to the latest time. Indeed, for the configurations with $S_{k_1}$ and $S_{k_2}$ the conditional probability $P_{k_1 | k_2}$ (information about the configuration $k_2$ is known) is related to the reverse conditional probability $P_{k_2 | k_1}$ according to the Bayes' theorem \cite{Bayes},
\begin{equation}
  P_{k_2}P_{k_1 | k_2} = P_{k_1}P_{k_2 | k_1} ~.
\end{equation}
In other words, one of the two conditional probabilities, but not both, can be given by the maximum entropy distribution. Note that the Bayes law expressed in terms of $I$ and $S$, which are related to the probabilities via exponents, is equivalent to (\ref{H=C}), or, according to (\ref{I+S+H}),
\begin{equation} \label{I+S=C}
  I + S = const ~,
\end{equation}
and shows that for the quasi-isolated systems the entropic and physical times are correlated.

Note that for a quantum system, by specific measurements which reduce marginal entropies in $H$, it is possible to achieve the state with
\begin{equation}
  I + H \approx const ~.
\end{equation}
This situation with
\begin{equation}
  S \approx const
\end{equation}
corresponds to the quantum Zeno effect when one can freeze the evolution of the quantum system by measuring it frequently enough \cite{Zeno}.

Since the law of evolution of $S$ may be considered as the definition of time $t$, one can introduce the quantity (energy) which is conserved during this evolution,
\begin{equation}
  E = \frac At ~.
\end{equation}
The energy $E$ can be assumed to be proportional to some universal parameter of the world ensemble that characterizes the property of constant entanglement (\ref{H=C}) for different 'classical' systems and which can have the dimension of the speed to bound the Lieb-Robinson velocities of entanglement propagation \cite{Lieb}. Then, in our model there naturally emerges the concept of mass,
\begin{equation}
  m = \frac {E}{c^2} ~,
\end{equation}
which characterizes the quasi-isolated system itself.


\section{Physical space}

The Relativity Principle we connect with the ambiguity of definition of three different types of entropy. For any physical system one can exchange constant additive terms of $I$, $S$ and $H$ with the rest of the universe, what does not affect dynamics of this system. Indeed, (i) the system can be situated in any 'place' of world ensemble (uncertainty of information), (ii) can have any 'speed' with respect to the universe (thermodynamic uncertainty due to 'heat' supply to the system) and (iii) any particle of the system can be replaced by analogous particle of the world ensemble (entanglement uncertainty). These ambiguities in definitions of the entropies, due to our assumption (\ref{S=A/h}), would exhibit necessity in three distinct parameters to define transformations of the classical action,
\begin{equation}\label{A}
  A = Et ~.
\end{equation}
On the level of action this 'entropic symmetry' can be parameterized by three spatial coordinates having the same character. For example, one can modify the classical action by adding of three extra terms,
\begin{equation}\label{A'}
  A' = A + mv^ix_i~, ~~~~~ (i = 1,2,3)
\end{equation}
which are described by $m$ and average Lieb-Robinson velocities, $v^i$, and introduce the 'spatial' dimensions, $x^i$, in this manner. Then based on the relation (\ref{S=A/h}) we can state that the less the probability of action transfer between two objects is, the further apart in this emergent space they are.

One can speculate that connected with $I$ dimension is subjective, while two other coordinates are objective, what can explain validity of the holography principle in field theory.

The symmetry of the action with extra terms (\ref{A'}) to be reduced to the old form (\ref{A}) (but with different energy) represents the Relativity Principle in emergent 4-dimensional 'space-time' of our model. In terms of the unit of the actions, $\hbar$, this symmetry can be written as de Broglie's phase harmony, which leads to the Bohr-Sommerfeld quantization condition in old quantum mechanics.


\section{Inertial frames and gravitational potential}

The Equivalence Principle also appears to be encoded in the properties of quantum entropy. Any sub-system of finite world ensemble which can be treated as isolated (external interactions are ignored) should have constant $S$ and the condition (\ref{I+S+H}) gives,
\begin{equation}
  dI = -dH ~.
\end{equation}
This property of the entanglement entropy that the marginal entropies of pure bi-partite states are equal, while the entropy of the overall universe remains zero \cite{Wilde}, means that for any physical system there exists the special state (inertial frame) in which information about the rest of the universe looks similar.

For a system with some action $A$ (or with the energy $E$) the relation (\ref{I+S+H}) takes the form:
\begin{equation}
  dI = - dS - dH = - \frac {dE}{T} - dH ~,
\end{equation}
where $T$ is the temperature of local environment. Then
\begin{equation}
  \frac {dE}{T} = \frac {dA}{\hbar} - dH = \left(1 - \frac {\phi}{c^2} \right) \frac {dA}{\hbar} ~,
\end{equation}
where we introduced the 'gravitational potential',
\begin{equation}
 \phi = \hbar c^2 \frac {dH}{dA} ~.
\end{equation}
The case with constant entanglement, $dH \to 0$, i.e. $\phi \to 0$, corresponds to the description of the universe by Minkowski space-time.


\section{The equilibrium condition}

The Minimum Action and the Maximum Entropy principles, used in physics and information theory, can be unified in a single dynamical equilibrium condition of our model.

Let us consider a physical system as a superposition of all its $k$ configurations, not only the most probable ones with the highest entropy. According to (\ref{S=A/h}), for the thermodynamic entropy we write
\begin{equation} \label{S=sum}
  S \sim \frac 1\hbar \sum_k P_k A_k = \frac 1\hbar \langle A \rangle~,
\end{equation}
where $A_k$ are the actions of the configurations and $\langle A \rangle$ is the expected action of the system. So (\ref{S=sum}) relates the thermodynamic entropy to the number of possible arrangements of $\hbar$ in $\langle A \rangle$. In the states close to the thermodynamic equilibrium (i.e. having the largest entropies) the transfers of elementary action quanta should be minimal and the probabilities can be written as
\begin{equation} \label{P=e^A}
  P_k = \frac 1Z e^{-A_k/\hbar} ~,
\end{equation}
where
\begin{equation}
  Z = \sum_k e^{-A_k/\hbar}
\end{equation}
is the partition function. Then
\begin{equation}
  A_k = - \hbar \left(\ln Z + \ln P_k \right)
\end{equation}
and (\ref{S=sum}) takes the Shannon's form (\ref{I}) up to the additive term $\ln Z$.

The variation of the action expectation over all configurations of the physical system, $\delta \langle A \rangle$, in general is not equal to $\langle \delta A \rangle$. To amend this incompleteness of optimization one can consider the so called stochastic action principle \cite{Wang, Bad},
\begin{equation} \label{bar-dA}
 \langle \delta A \rangle = \sum_k P_k \delta A_k = \delta \langle A \rangle - \hbar \delta S = 0~.
\end{equation}
The last step in this formula follows from (\ref{P=e^A}) and the trivial decomposition
\begin{equation}
  \delta \langle A \rangle = \delta \sum_k P_k A_k = \langle \delta A \rangle + \sum_k A_k \delta P_k ~.
\end{equation}
If we introduce a temperature and a generalized heat, such that
\begin{equation}
  \delta S = \frac {\delta Q}{T} ~,
\end{equation}
equation (\ref{bar-dA}) mimics the first law of thermodynamics. Note that the condition (\ref{bar-dA}) is equivalent to a maximization of entropy
\begin{equation}
  \delta S = \frac 1\hbar \left( \delta \langle A \rangle - \langle \delta A \rangle\right) ~,
\end{equation}
i.e. the maximum entropy principle in thermodynamics \cite{Jay} and the least action principle in field theory can be unified into a single principle.

For the systems which obey (\ref{I+S=C}) the entropy in (\ref{bar-dA}) is given by the Shannon's formula (\ref{I}). In general, the entanglement entropy $H$ is not constant, that can be understood as the dissipation of information in the environment of a quasi-isolated system. Then, as usual in mathematical descriptions of dissipative processes, the expression for $S$ should be modified by addition of some complex terms:
\begin{equation}
  S_{tot} = S + i \sum_n p_n H_n ~,
\end{equation}
where $p_n$ are the probabilities of interaction of the system with other $n$ subsystems of the universe. Using the analogies with $P_k$, which are related with $A_k$ via the exponentials (\ref{P=e^A}), we can assume the Feynman's form for the probability-action relation,
\begin{equation}
  p_n \sim e^{i A_n/\hbar} ~.
\end{equation}
Then we obtain the real value for the second term in $S_{tot}$, which, like the first term, also takes the form of (\ref{I}). Therefore, the total probability, which describes the behavior of particles of the selected system, can be written as the product of $P_k$ and $p_n$ and it imitates the structure of the wave function in quantum mechanics. If the action of the system is given by the form of an ideal fluid, then it can be cast into the action for the Schr\"{o}dinger equation \cite{Gog-3, Cat}. This suggests that the field variables and the space-time structure for quantum ensembles have the same origin and are not to be defined separately.

Finally the main dynamical equation of our model (\ref{bar-dA}) takes the form:
\begin{equation} \label{dyn-eq}
  \delta \left(S_{tot} - \frac 1\hbar \langle A \rangle \right) = 0~,
\end{equation}
and, in addition to the classical action $A$, contains different types of entropies in $S_{tot}$. It is known that the Einstein's equations can be obtained from the relation similar to (\ref{dyn-eq}), which is equivalent to the first law of thermodynamics \cite{Jac, Ver}. For the recent progress in deriving the gravitational dynamics from the entanglement entropy see \cite{Grav-1, Grav-2, Grav-3, Grav-4}, and some attempts to apply the entanglement in cosmology are discussed in \cite{Cos-1, Cos-2, Cos-3, Cos-4}.

Note that the equilibrium condition of a physical system and the world ensemble in the form (\ref{dyn-eq}) can solve the old problem with the non-locality of the standard least action principle, since this difficulty of classical formulation (a physical system knows itself how to behave and what is the geodesic trajectory \cite{Poincare}; in nature there exist meaningless intentions and the present depends on the future events \cite{non-local-1, non-local-2}) originates in the ignorance of quantum-information properties of physical systems.


\section{Discussion}

The main hypothesis of this paper is that the universe can be described as a single entangled ensemble of quantum particles which obey the Zero Entropy Principle (\ref{I+S+H}). The consequence of this assumption for a classical subsystem with constant entanglement entropy, and which also obey the principle of maximum statistical entropy ($S$ is maximal and thus $I$, due to (\ref{I+S=C}), is minimal), is that an observer has maximal information about real trajectories of the system and can track its motion. For a 'quantum system' the information entropy, $I$, is large and he is not certain about its physical characteristics. In other words, an observer must be considered as part of the experiment, gain of information by him reduces uncertainty of measurements and vice versa. One observable evidence of this statement of the model is that enough large number of observers, by the concentration on some prior, in principle are able to change the probability distribution for some random variable.


\section*{Acknowledgment:}

This research was partially supported by the Shota Rustaveli National Science Foundation grant ST $09\_798\_4-100$.



\begin{thebibliography}{99}

\bibitem{Car} R. Carroll,
             {\it On the Emergence Theme of Physics}
             (World Scientific, Singapore 2010).

\bibitem{deBr} L. de Broglie,
              Ann. Inst. Poincare {\bf A 1} (1964) 1.

\bibitem{Sakh} A.D. Sakharov,
              Sov. Phys. Dokl. {\bf 12} (1968) 1040
              (Gen. Rel. Grav. {\bf 32} (2000) 365).

\bibitem{Bek} J.D. Bekenstein,
             Phys. Rev. {\bf D 7} (1973) 2333.

\bibitem{BCH} J.M. Bardeen, B. Carter and S.W. Hawking,
             Comm. Math. Phys. {\bf 31} (1973) 161.

\bibitem{Haw-1} S.W. Hawking,
              Comm. Math. Phys. {\bf 43} (1975) 199;
              Comm. Math. Phys. {\bf 46} (1976) 206.

\bibitem{Rov-2} A. Connes and C. Rovelli,
               Class. Quant. Grav. {\bf 11} (1994) 2899, arXiv: gr-qc/9406019.

\bibitem{Rov-3} C. Rovelli and M. Smerlak,
               Class. Quant. Grav. {\bf 28} (2011) 075007, arXiv: 1005.2985 [gr-qc].

\bibitem{Jac} T. Jacobson,
             Phys. Rev. Lett. {\bf 75} (1995) 1260, arXiv: 9504004 [gr-qc].

\bibitem{Pan} T. Padmanabhan,
             Class. Quant. Grav. {\bf 21} (2004) 4485, arXiv: gr-qc/0308070;
             Rep. Prog. Phys. {\bf 73} (2010) 046901, arXiv: 0911.5004 [gr-qc];
             J. Phys. Conf. Ser. {\bf 306} (2011) 012001, arXiv: 1012.4476 [gr-qc];
             Entropy {\bf 17} (2015) 7420, arXiv: 1508.06286 [gr-qc].

\bibitem{Ver} E.P. Verlinde,
             JHEP {\bf 1104} (2011) 029, arXiv: 1001.0785 [hep-th].

\bibitem{Stat-1} F. Reif,
               {\it Fundamentals of Statistical and Thermal Physics} (McGraw-Hill, New York 1965).

\bibitem{Stat-2} J.R. Waldram,
                {\it The Theory of Thermodynamics} (Cambridge University Press, Cambridge 1985).

\bibitem{Shan} C.E. Shannon,
              Bell Sys. Tech. J. {\bf 27} (1948) 379.

\bibitem{Ent-1} I. Bengtsson and K. \v{Z}yczkowski,
             {\it Geometry of Quantum States} (Cambridge University Press, Cambridge 2007).

\bibitem{Ent-2} R. Horodecki, P. Horodecki, M. Horodecki and K. Horodecki,
             Rev. Mod. Phys. {\bf 81} (2009) 865, arXiv: quant-ph/0702225.

\bibitem{Ent-3}J. Eisert, M. Cramer and M. B. Plenio,
             Rev. Mod. Phys. {\bf 82} (2010) 277, arXiv: 0808.3773 [quant-ph].

\bibitem{Phas-Tr-1} S.L. Sondhi, S.M. Girvin, J.P. Carini and D. Shahar,
                   Rev. Mod. Phys. {\bf 69} (1997) 315, arXiv: cond-mat/9609279.

\bibitem{Phas-Tr-2} S. Sachdev,
                   {\it Quantum Phase Transitions} (Cambridge University Press, Cambridge 2011).

\bibitem{BH-1} L. Bombelli, R.K. Koul, J. Lee and R.D. Sorkin,
              Phys. Rev. {\bf D 34} (1986) 373.

\bibitem{BH-2} M. Srednicki,
              Phys. Rev. Lett. {\bf 71} (1993) 666, arXiv: hep-th/9303048.

\bibitem{BH-3} S.N. Solodukhin,
              Living Rev. Rel. {\bf 14} (2011) 8, arXiv: 1104.3712 [hep-th].

\bibitem{H=S-1} J.M. Deutsch, H. Li and A. Sharma,
               Phys. Rev. {\bf E 87} (2013) 042135, arXiv: 1202.2403 [quant-ph].

\bibitem{H=S-2} G. Chirco, H.M. Haggard, A. Riello and C. Rovelli,
               Phys. Rev. {\bf D 90} (2014) 044044, arXiv: 1401.5262 [gr-qc].

\bibitem{S-div-1} C.G. Callan (Jr.) and F. Wilczek,
                 Phys. Lett. {\bf B 333} (1994) 55, arXiv: hep-th/9401072.

\bibitem{S-div-2} T. Padmanabhan,
                 Phys. Rev. {\bf D 82} (2010) 124025, arXiv: 1007.5066 [gr-qc].

\bibitem{S-div-3} D. Nesterov and S.N. Solodukhin,
                 JHEP {\bf 1009} (2010) 041, arXiv: 1008.0777  [hep-th].

\bibitem{Schl} F. Schl\"{o}gl,
              Am. J. Phys. {\bf 78} (2010) 7.

\bibitem{Wilde} M.M. Wilde,
               {\it Quantum Information Theory}
               (Cambridge University Press, Cambridge 2013), arXiv: 1106.1445 [quant-ph].

\bibitem{I-cons} R. Landauer,
                Phys. Today {\bf 44} (1991) 23.

\bibitem{I-clon} W.K. Wootters and W.H. Zurek,
                Nature {\bf 299} (1982) 802; Phys. Today {\bf 62} (2009) 76.

\bibitem{I-del} A.K. Pati and S.L. Braunstein,
               Nature 404 (2000) 164, arXiv: quant-ph/9911090.

\bibitem{Teleport} C.H. Bennett, G. Brassard, C. Cr\'{e}peau, R. Jozsa, A. Peres, and W. Wootters,
                  Phys. Rev. Lett. {\bf 70} (1993) 1895.

\bibitem{Neg-S-1} G. Vidal and R. F. Werner,
                 Phys. Rev. {\bf A 65} (2002) 032314, arXiv: quant-ph/0102117.

\bibitem{Neg-S-2} M. B. Plenio,
                 Phys. Rev. Lett. {\bf 95} (2005) 090503, arXiv: quant-ph/0505071.

\bibitem{Neg-S-3} R. Horodecki, P. Horodecki, M. Horodecki and K. Horodecki,
                 Rev. Mod. Phys. {\bf 81} (2009) 865, arXiv: quant-ph/0702225.

\bibitem{Neg-S-4} D. Singleton, E.C. Vagenas, T. Zhu and J.-R. Ren,
                 JHEP {\bf 1008} (2010) 089; Erratum: JHEP {\bf 1101} (2011) 021, arXiv: 1005.3778 [gr-qc].

\bibitem{Neg-S-5} P. Calabrese, J. Cardy and E. Tonni,
                 Phys. Rev. Lett. {\bf 109} (2012) 130502, arXiv: 1206.3092 [cond-mat.stat-mech].

\bibitem{Neg-S-6} D. Singleton, E.C. Vagenas and T. Zhu,
                 JHEP {\bf 1405} (2014) 074, arXiv: 1311.2015 [gr-qc].

\bibitem{Neg-S-7} M. Rangamani and M. Rota,
                 JHEP {\bf 1410} (2014) 60, arXiv: 1406.6989 [hep-th].

\bibitem{Neg-S-8} S. Kanno, J.P. Shock and J. Sodaz,
                 JCAP {\bf 1503} (2015) 015, arXiv: 1412.2838 [hep-th].

\bibitem{Ent-S-1} P. Calabrese and J. L. Cardy,
                 J. Stat. Mech. {\bf 0406} (2004) P06002, hep-th/0405152;
                 J. Phys. {\bf A 42} (2009) 504005, arXiv: 0905.4013 [cond-mat.stat-mech].

\bibitem{Ent-S-2} C. Holzhey, F. Larsen and F. Wilczek,
                 Nucl. Phys. {\bf B 424} (1994) 443, hep-th/9403108.

\bibitem{HOW} M. Horodecki, J. Oppenheim and A. Winter,
             Nature {\bf 436} (2005) 673, arXiv: quant-ph/0505062.

\bibitem{RARDV} L. del Rio, J. Aberg, R. Renner, O. Dahlsten, and V. Vedral,
               Nature {\bf 474} (2011) 61, arXiv: 1009.1630 [quant-ph].

\bibitem{Song} D. Song,
              Int. J. Theor. Phys. {\bf 53} (2014) 1369, arXiv: 1302.6141 [physics.gen-ph].

\bibitem{Cerf-Adami} N.J. Cerf and C. Adami,
                    Phys. Rev. Lett. {\bf 79} (1997) 5194, arXiv: quant-ph/9512022.

\bibitem{Zei-1} A. Zeilinger,
               Found. Phys. {\bf 29} (1999) 631;
               Rev. Mod. Phys. {\bf 71} (1999) S288.

\bibitem{Zei-2} C. Brukner and A. Zeilinger,
               Phys. Rev. Lett. {\bf 83} (1999) 3354, arXiv: quant-ph/0005084;
               Phys. Rev. {\bf A 63} (2001) 022113, arXiv: quant-ph/0006087;
               in {\it Time, Quantum, Information}, eds. L. Castell and O. Ischebeck (Springer, Berlun 2003), arXiv: quant-ph/0212084.

\bibitem{Penrose} R. Penrose
                 in {\it General Relativity: An Einstein Centenary Survey}, eds. S.W. Hawking and W. Israel (Cambridge Univ. Press, Cambridge 1979);
                 in {\it 300 Years of Gravitation}, eds. S.W. Hawking and W. Israel (Cambridge Univ. Press, Cambridge 1987);
                 {\it The Big Bang and its Thermodynamic Legacy in Road to Reality} (Vintage Books, London 2004).

\bibitem{Lin-Ega} C.H. Lineweaver  and C.A. Egan,
                 Phys. Life Rev. {\bf 5} (2008) 225;
                 Astrophys. J. {\bf 710} (2010) 1825, arXiv: 0909.3983 [astro-ph.CO].

\bibitem{Hwan} W.-Y. Hwang,
              Nat. Sci. {\bf 6} (2014) 540, arXiv: 0806.2749 [quant-ph].

\bibitem{Araki-Lieb} H. Araki and E.H. Lieb,
                    Commun. Math. Phys. {\bf 18} (1970) 160.

\bibitem{Info-1} L.B. Kish,
                Fluct. Noise Lett. {\bf 7} (2007) C51, arXiv: 0711.1197 [physics.gen-ph].

\bibitem{Info-2} L.B. Kish, C.G. Granqvist,
                Proc. IEEE {\bf 9} (2013) 1895, arXiv: 1309.7889 [physics.class-ph].

\bibitem{Info-3} L. Herrera,
                Fluc. Noise Lett. {\bf 13} (2014) 1450002, arXiv: 1403.4511 [gr-qc].

\bibitem{Land-1} R. Landauer,
                IBM J. Res. Develop. {\bf 5} (1961) 183.

\bibitem{Land-2} A. B\'{e}rut, A. Arakelyan, A. Petrosyan, S. Ciliberto, R. Dillenschneider and E. Lutz,
                Nature {\bf 483} (2012) 187.

\bibitem{Bril} L. Brillouin,
              J. Appl. Phys. {\bf 24} (1953) 1152;
              {\it Science and Information Theory} (Academic, New York 1962);
              {\it Scientific Uncertainty and Information} (Academic, New York 1964).

\bibitem{Unr} W.G. Unruh,
             Phil. Trans. R. Soc. {\bf A 370} (2012) 4454, arXiv: 1205.6750 [quant-ph].

\bibitem{JMK} R.H. Jonsson, E. Martin-Martinez and A. Kempf,
             Phys. Rev. Lett. {\bf 114} (2015) 110505, arXiv: 1405.3988 [quant-ph].

\bibitem{Ann} A. Annila,
              Entropy {\bf 12} (2010) 2333, arXiv: 1005.3854 [physics.gen-ph].

\bibitem{Fey} R.P. Feynman, F.B. Morinigo and G. Wagner,
             {\it Feynman Lectures on Gravitation} (Addison-Wesley, Reading 1995).

\bibitem{Haw-2} S. Hawking,
              {\it A Brief History of Time} (Bantam, Toronto 1988).

\bibitem{Gog-1} M. Gogberashvili,
             Eur. Phys. J. {\bf C 54} (2008) 671, arXiv: 0707.4308 [hep-th];
             Eur. Phys. J. {\bf C 63} (2009) 317, arXiv: 0807.2439 [gr-qc].

\bibitem{Gog-2} M. Gogberashvili and I. Kanatchikov,
               Int. J. Theor. Phys. {\bf 51} (2012) 985, arXiv: 1012.5914 [physics.gen-ph].

\bibitem{Gog-3} M. Gogberashvili,
              Int. J. Theor. Phys. {\bf 50} (2011) 2391, arXiv: 1008.2544 [gr-qc].

\bibitem{Jay} E.T. Jaynes,
             Phys. Rev. {\bf 106} (1957) 620;
             Amer. J. Phys. {\bf 33} (1965) 391;
             in {\it Papers on Probability, Statistics and Statistical Physics}, ed. R.D. Rosenkrantz (Reidel, Dordrecht 1983);
             in {\it Maximum Entropy and Bayesian Methods in Inverse Problems}, eds. C.R. Smith and W.T. Grandy Jr. (Reidel, Dordrecht 1985).

\bibitem{Bayes} A. Stuart and K. Ord,
              {\it Kendall's Advanced Theory of Statistics, Volume I: Distribution Theory}, Sec. 8.7 (Arnold, London 1994).

\bibitem{Zeno} P. Facchi and S. Pascazio,
              J. Phys. {\bf A 41} (2008) 493001, arXiv: 0903.3297 [math-ph].

\bibitem{Lieb} E.H. Lieb and D.W. Robinson,
              Commun. Math. Phys. {\bf 28} (1972) 251.

\bibitem{Wang} Q.A. Wang,
             Chaos, Solit. Fract. {\bf 23} (2004) 1253, arXiv: cond-mat/0405373;
             ibid. {\bf 26} (2005) 1045, arXiv: cond-mat/0412360;
             Astrophys. Space Sci. {\bf 305} (2006) 273, arXiv: cond-mat/0312329.

\bibitem{Bad} J.P. Badiali,
             J. Phys. {\bf A 39} (2006) 7175, arXiv: gr-qc/0505050.

\bibitem{Cat} A. Caticha,
             J. Phys. {\bf A 44} (2011) 225303, arXiv: 1005.2357 [quant-ph].

\bibitem{Grav-1} N. Lashkari, M. B. McDermott and M. Van Raamsdonk,
                JHEP {\bf 1404} (2014) 195, arXiv: 1308.3716 [hep-th].

\bibitem{Grav-2} M. Nozaki, T. Numasawa, A. Prudenziati and T. Takayanagi,
                Phys. Rev. {\bf D 88} (2013) 026012, arXiv: 1304.7100 [hep-th].

\bibitem{Grav-3} J. Bhattacharya and T. Takayanagi,
                JHEP {\bf 2013} (2013) 219, arXiv: 1308.3792 [hep-th].

\bibitem{Grav-4} T. Faulkner, M. Guica, T. Hartman, R. C. Myers and M. Van Raamsdonk,
                JHEP {\bf 2014} (2014) 51, arXiv: 1312.7856 [hep-th].

\bibitem{Cos-1} J.L. Ball, I. Fuentes-Schuller and F. P. Schuller,
               Phys. Lett. {\bf A 359} (2006) 550, arXiv: quant-ph/0506113.

\bibitem{Cos-2} G.L. Ver Steeg and N.C. Menicucci,
               Phys. Rev. {\bf D 79} (2009) 044027, arXiv: 0711.3066 [quant-ph].

\bibitem{Cos-3} Y. Nambu and Y. Ohsumi,
               Phys. Rev. {\bf D 84} (2011) 044028, arXiv: 1105.5212 [gr-qc].

\bibitem{Cos-4} A. Albrecht, N. Bolis and R. Holman,
               JHEP {\bf 2014} (2014) 93, arXiv: 1408.6859 [hep-th].

\bibitem{Poincare} H. Poincar\'{e},
                  {\it La Science et l'Hypoth\`{e}se} (Flammarion, Paris 1902).

\bibitem{non-local-1} H. Hertz,
                     {\it The Principles of Mechanics, Presented in a New Form} (Dover, New York 1956).

\bibitem{non-local-2} V. Berdichevsky,
                    {\it Variational Principles of Continuum Mechanics} (Springer, Berlin 2009).

\end{thebibliography}
\end{document}